ARTICLE TEMPLATE

# Subgroup Discovery in MOOCs: A Big Data Application for Describing Different Types of Learners

J. M. Luna[a] and H. M. Fardoun[b] and F. Padillo[a] and C. Romero[a] and S. Ventura[a,b]

[a]Department of Computer Science and Numerical Analysis, University of Cordoba, Cordoba, Spain
[b]Faculty of Computing and Information Technology, King Abdulaziz University, Saudi Arabia

**ABSTRACT**
The aim of this paper is to categorize and describe different types of learners in massive open online courses (MOOCs) by means of a subgroup discovery approach based on MapReduce. The final objective is to discover IF-THEN rules that appear in different MOOCs. The proposed subgroup discovery approach, which is an extension of the well-known FP-Growth algorithm, considers emerging parallel methodologies like MapReduce to be able to cope with extremely large datasets. As an additional feature, the proposal includes a threshold value to denote the number of courses that each discovered rule should satisfy. A post-processing step is also included so redundant subgroups can be removed. The experimental stage is carried out by considering de-identified data from the first year of 16 MITx and HarvardX courses on the edX platform. Experimental results demonstrate that the proposed MapReduce approach outperforms traditional sequential subgroup discovery approaches, achieving a runtime that is almost constant for different courses. Additionally, thanks to the final post-processing step, only interesting and not-redundant rules are discovered, hence reducing the number of subgroups in one or two orders of magnitude. Finally, the discovered subgroups are easily used by courses' instructors not only for descriptive purposes but also for additional tasks such as recommendation or personalization.

## 1. Introduction

Over the last few years, there has been an increasing interest in massive open online courses (MOOCs), which have changed the online education. MOOCs have emerged as a really popular learning model in which students may interact with other students, professors or even teaching assistants. MOOCs include an enormous number of learners with largely unknown diversity in terms of ability, motivation as well as real goals. With no cost to entry or exit, learners in MOOCs include a wide range of backgrounds, intentions and personal or technical constraints to participate. Anyone having Internet access might register for a course and no matter whether it was by simple curiosity or purely by accident (Liyanagunawardena, 2014). As a result, it turns necessary to

CONTACT S. Ventura. Email: sventura@uco.es

analyse the huge volumes of data that are daily generated in MOOCs in order to describe, understand or predict, in a near future, features such as the students' profile, engagement and behaviour within the courses (Romero, 2017). In this way, currently, there is an increasing interest in discovering different types of learners in MOOCs based on the obtained insights (Anderson, 2014; Douglas, 2016; Ferguson, 2015; Hill, 2013; Kizilcec, 2013; Sharma, 2015). The analysis and categorization of insights that appear in different MOOCs is a very important open research question in learning analytics (Ferguson, 2015) and few works have addressed this alluring issue (Bohannon, 2014). A good example of discovering features that appear in different MOOCs was addressed by *Fergunson and Clow* (Ferguson, 2015), who proposed a clustering algorithm to a series of courses already analysed by *Kizilcec et al.* (Kizilcec, 2013). As a result, only two of the four discovered clusters were already identified by *Kizilcec et al.* (Kizilcec, 2013). In a more recent and different approach, *Andres et al.* (Andres, 2017) proposed a production rule system for representing existing findings in an understandable fashion for researchers and practitioners. After reading 68 published papers about MOOCs, the findings were manually transformed into a total of 21 IF-THEN rules.

Continuing with the same idea of *Andres et al.* (Andres, 2017), in which IF-THEN rules were used to represent students' behaviour, the aim of this paper is to propose a subgroup discovery (SD) algorithm that automatically discovers all these rules. By means of this supervised descriptive approach, the process of describing different types or categories of students in MOOCs can be eased. In this regard, data from several MOOC courses are analysed with the aim of discovering rules that appears in multiple courses. The main objetive is therefore to discover the most interesting (not redundant and accurate) IF-THEN rules for describing different types or categories of learners on MOOCs. The proposed algorithm is based on traditional SD approaches but using emerging parallel technologies for Big Data like MapReduce (Dean, 2008), which has proved to obtain excellent results on dealing with massive datasets. It is important to highlight that categorizing and describing types of learners are really important tasks that turn into a time consuming and arduous process, specially for huge amount of data. This aspect is specially important since MOOCs usually gather high dimensional data that hamper the process of extracting useful and unexpected insights through traditional machine learning approaches. More specifically, the proposal was implemented on Apache Spark since it has proved to be the best solution to model iterative algorithms (Zaharia, 2010) with MapReduce.

The rest of the paper is structured as follows. Section 2 presents the most relevant definitions and related works. Section 3 describes the proposed approach; Section 4 presents the experimental analysis; Finally, some concluding remarks are outlined in Section 5.

## 2. Background

In this section, the problem of categorizing students in MOOCs is first introduced. Then, the subgroup discovery task is described as well as its application to the educational field. Finally, the foundations of MapReduce and Apache Spark are presented.

### *2.1. Categorizing learners in MOOCs*

MOOC learners generally sing up for the course with a wide variety of objectives (Kizilcec, 2015). Given the heterogeneity of the learners, it would be remiss to make prior assumptions about the appropriate characteristics or behaviours in which learners might be categorized(Kizilcec, 2013). Few studies, however, are based on identifying the types of learners who enroll in a MOOC course (Woodgate, 2015). One of these studies was proposed by *Phil Hill* (Hill, 2013), in which students in a MOOC were categorized into Lurkers (those who only enroll in the course), Active (fully engaged with the course material, quizzes and forums), Passive (they only consume the content but they do not actively participate in forums) and Drop-ins (they are active students that only work on a small part of the course). Some of these categories were, additionally, revised and divided: Lukers into No-shows (they register for the course but they never log in) and Observers (they login and may read contents or browse discussions, but they do not usually take any form of assessment).

*Anderson et al.* (Anderson, 2014) formalized a taxonomy of individual behaviour in a MOOC based on the relative frequency of certain activities undertaken by each student. Authors used these trace data of several Coursera courses to discover the next styles of engagement: Viewers (they generally watch video-lectures, but they perform only few assignments), Solvers (they submit assignments for a grade, but they watch few video-lectures), All-rounders (there is a balance between video-lectures watched and number of submissions), Collectors (they generally download video-lectures and and no submissions are generally done), and Bystanders (they have a very low activity in the course).

*Sharma et al.* (Sharma, 2015) proposed a hierarchical scheme to categorize students into two major groups: Active Students and Viewers. Active Students are those students who actively participate in the course, i.e. they take part in the assessment processes. Viewers, on the contrary, are those students who just watch the videos from the course. These students are further divided based on two additional factors. First, whether they assess their learning by means of non-mandatory quizzes. Second, the amount of videos they watch. Considering the first factor, a student can be subdivided into Active Viewers and Passive Viewers. As for the second factor, students can be subdivided into Wiki Viewers (students watching less than 10% of the videos), Dropouts (students watching between 10% and 70% of the videos) and Completers (students that watch more than 70% of the videos).

*Kizilcec et al.* (Kizilcec, 2013) analysed patterns of engagement and disengagement from three different MOOCs available at the Coursera platform. They used a clustering analysis based on learners' patterns of interaction with video lectures and assessments. Four groups were obtained: Completing learners (they usually complete most of the assessments), Auditing learners (they tend to watch most of the videos but they do not complete all the assessments), Disengaging learners (they generally complete all the assessments at the beginning of the course but they dradually reduce their activity), and Sampling learners (they simply explore some videos).

*Fergunson and Clow* (Ferguson, 2015) analysed engagement patterns on four FutureLearn MOOCs using the *k*-means clustering algorithm. They discovered seven distinct patterns of engagement: (1) Samplers, which were learners who briefly visited the course; (2) Strong Starters, those that completed the first assessment of the course but they dropped out; (3) Returners, learners that completed the assessment in the first week, returned in the second week, and then dropped out; (4) Mid-way Dropouts, those who completed three or four assessments but they finally dropped out; (5) Nearly There, learners that consistently completed assessments but they dropped out just before the end of the course; (6) Late Completers, those who completed the final assessment, submitted most of the other assessments, but they were either late or missed some out; and, finally (7) Keen Completers, learners who completed the course

| Algorithm | Article | #Clusters | Label for each cluster |
|---|---|---|---|
| k-means | (Kizilcec, 2013) | 4 | Completing, Auditing, Disengaging, Sampling |
| | (Ferguson, 2015) | 7 | Samplers, Strong Starters, Returners, Mid-way Dropouts, Nearly There, Late Completers, Keen Completers learners |
| | (Rodriguez, 2016) | 3 | Engaged, Sporadic, Disengaged |
| | (Kovanovic, 2016) | 5 | Enrollees, Low Engagement, Videos, VideosQuizzes, Social |
| | (Khalil, 2017) | 4 | Dropout, Perfect Students, Gaming the System, Social |
| k-means++ | (Douglas, 2016) | 5 | Fully Engaged, Consistent Viewers, Two-Week Engaged, One-Week Engaged, Sporadic Learners |

**Table 1.** Comparison of studies based on clustering for categorizing MOOC students.

diligently, engaging actively throughout.

*Douglas et al.* (Douglas, 2016) applied a clustering algorithm (*k*-means++) on learner clickstream patterns. They discovered five types of MOOC learners that were labelled according to how they generally used the materials: Fully Engaged Learners (they had regular access to the study materials and completed most of the quizzes and exams), Consistent Viewers (they actively accessed to the content of the course but they did not regularly access to the course assessment materials), Two-Week Engaged Learners (they accessed to most of the materials during the first two weeks of the course, but they reduced their activities in subsequent weeks), One-Week Engaged Learners (they actively accessed to the materials during the first week but, afterwards, they reduced the participation), Sporadic Learners (there was a random access to the materials).

According to the aforementioned studies, most of the existing works describe categorization schemes depending on student's engagement. Some of these works are based on clustering approaches to automatically discover different types of learners in MOOCs. A summary table (see Table 1) in which different authors propose different algorithms and type of learners depending on the obtained results. It is important to highlight that the goal of clustering in this problem was to is to identify groups of transactions (learners in this case) that equally behave. Thus, these solutions cannot be compared to the subgroup discovery task since their aims are completely different. Subgroup discovery aims at describing set of learners (already predefined), whereas clustering aims at discovering how many set of learners can be formed in a dataset due to they share some features.

Although a varied set of learners have been identified by different authors through clustering algorithms, in most of the cases, these types are very similar or even the same (different names are considered). Additionally, most of the works describe only three, four or five categories and just only one of the aforementioned research studies (Sharma, 2015) described the groups in a more general detail, considering also a second level or subtypes of learners.

### 2.2. Subgroup Discovery

Subgroup discovery (SD) is a supervised descriptive pattern mining (Ventura, 2018) technique that seeks relations among different variable values with respect to a class attribute. SD lies halfway between unsupervised (describing hidden structures by exploring unlabeled data) and unsupervised (obtaining an accurate classifier that predicts the correct output given an input) learning tasks. Unlike classification tasks, SD seeks to extract interesting subgroups (not an accurate classifier) for a given class or target attribute and the discovered subgroups do not necessarily cover all the examples for the class. Similarly, unlike clustering, where the goal is to describe unlabeled data structure, SD was conceived to form subgroups that similarly behave and, therefore, new input data could be classified according to these subgroups. Figure 1 shows the difference between the classification, clustering and SD tasks.

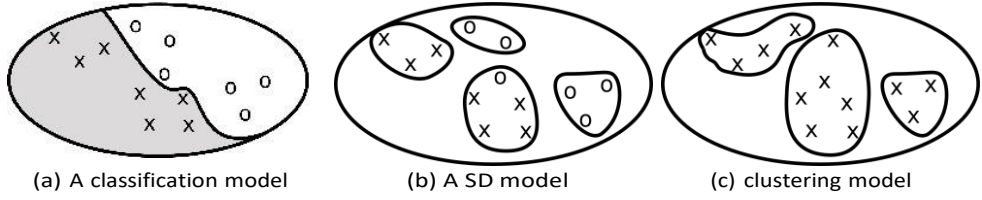

(a) A classification model     (b) A SD model     (c) clustering model

**Figure 1.** Difference between classification, clustering and subgroup discovery tasks

SD represents relationships by means of rules of the form $R \equiv ant \rightarrow t$, where *ant* is a set of independent variables known as subgroup, and *t* represents a value of the target or class variable. Aiming at quantifying how interesting the solutions are, *Herrera et al.* (Herrera, 2011) proposed to quantify the quality of the subgroups according to how many patterns are covered by each target (see Equation 1), that is, the number of transactions satisfied by both antecedent and target, with respect to the total number of examples covered by the target.

$$Support_{target}(R) = \frac{|\{ant \cup target \subseteq t, t \in DB\}|}{|\{target \subseteq t, t \in DB\}|} \quad (1)$$

*Herrera et al.* (Herrera, 2011) also considered the confidence measure (see Equation 2) as one of the most well-known and important measures since it quantifies the reliability of the subgroups.

$$Confidence(R) = \frac{|\{ant \cup target \subseteq t, t \in DB\}|}{|\{ant \subseteq t, t \in DB\}|} \quad (2)$$

SD has been successfully applied to different real world problems including education (Romero, 2009), describing relationships between the student's usage of the different activities and their final marks; as well as recommender systems such as the one proposed by *Luna et al.* (Luna, 2016) in which students' attitudes in high school were considered to recommend a specific degree to be studied in the future.

### *2.3. MapReduce*

MapReduce (Dean, 2008) is a recent framework for distributed computing, where algorithms are composed of two main stages: map and reduce. In the map phase, each mapper processes a subset of input data and produces a set of $\langle k, v \rangle$ pairs. Finally, the reducer takes this new list as input to produce the final values. Since its origins many implementations have been proposed (Lam, 2010; Zaharia, 2010).

MapReduce has been previously applied to the SD task in only few works but existing algorithms have demonstrated the usefulness of using this framework in the SD field (Padillo, 2017). As for educational data, there are some review works (Wassan, 2015) that described the potential usefullnes of parallel programming models like Hadoop MapReduce for accelerating the analysis of educational data. They showed that Hadoop can help in building scalable models in the field of education and may provide a better scope of improvement in the field of educational analytics. Far from the SD task, MapReduce have been applied to educational problems to process students' data activity in a Peer-to-peer Networked Setting (Santi, 2015). *Lasheng et al.* (Lasheng, 2017) considered Hadoop to perform a normalization of massive education data. *Raju et al.* (Raju, 2011) suggested a clustering algorithm called KMR to customize the E-Learning application. *Tajunisha and Anjali* (Tajunisha, 2015) used Mapreduce to predict students' performance. They proposed an extensional MapReduce Task Scheduling algorithm for Deadline constraints for improving classification accuracy even in the big data and reducing the time complexity.

## 3. Proposed Methodology

The proposed methodology for describing types of learners in MOOCs is based on the traditional three main stages of data mining (see Figure 2). First, a preprocessing procedure is carried out were MOOC data about students' information is gathered and structured. Second, a SD task adapted to MapReduce is performed. Finally, a post-processing step is responsible for filtering redundant rules as well as selecting the most interesting ones.

### 3.1. Pre-processing

In this first procedure, all the gathered information about students on different MOOC courses is pre-processed according to the following steps:

**Step 1.1** Attributes Selection. A list of variables or attributes with information about students in the MOOC is selected to be used as descriptors of the subgroups, that is, in the antecedent of the SD rules. In our case, a public dataset that directly provides all the students' information in aggregate records (each record represents a student) is considered. This dataset also provides the target (or targets) to be used in the SD rules. Thus, no attribute selection process is required in our analysis except for removing the users' id (unique value that identifies each student and, therefore, useless for the analysis).

**Step 1.2** Attributes Discretization. Any numerical attribute is transformed into a finite set of categories or discrete values. This transformation can be performed either in a manual manner of by considering some well-known discretization algorithms (Dougherty, 1995).

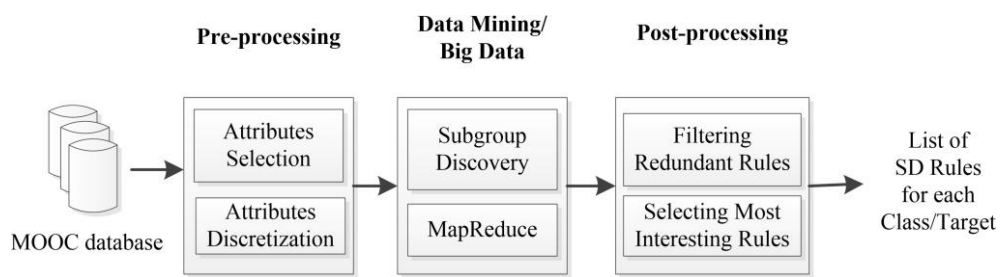

**Figure 2.** Proposed Approach for describing types of learners in MOOCs.

## 3.2. Data Mining

A specific SD algorithm called SD-DFPTRee (Subgroup Discovery using Distributed FP-Trees) (Padillo, 2017) is considered in the proposed methodology for discovering IF-THEN engagement rules over MOOCs. This algorithm is able to discover subgroups by using a special data structure known as FP-Tree in a distributed way, and it also includes some ideas of the parallel version of FP-Growth (Ventura, 2016). This algorithm was specifically designed to work with extremely huge datasets by considering the MapReduce framework. SD-DFPTRee includes the following steps:

**Step 2.1** The driver reads the dataset from disk, splitting it in successive parts and loaded in a Resilient Distributed Dataset (RDD). RDD represents a read-only collection of objects partitioned across a set of machines, enabling the dataset to be loaded in memory one time and read multiple times from executor process without having to load it in each iteration as Hadoop does.

**Step 2.2** A parallel counting is carried out (see Algorithm 1). In this step the frequency for each value of each attribute is calculated, resulting in a final list called F-list. These frequencies will be used to calculate the final quality measure for each subgroup.

**Step 2.3** F-list is sorted by frequency. This step is performed in a single computer in few seconds, since the size of F-list is enough smaller to be saved in main memory.

**Step 2.4** This final step is responsible for creating the FP-Trees. The mapper phase reads the RDD that was previously created in Step 2.2 and the sorted F-list obtained in Step 2.3. Each mapper produces a set of $(k, v)$ pairs, where $k$ is an item from F-list and $v$ is a item-dependent transaction. For each item within F-list, if it appears in a transaction then it locates its right-most appearance, say $L$, and outputs a key-value pair with the form $(item, instance[1]...instance_i[L])$. These steps are fully described in the function *mapperPFP* in Algorithm 2. A detailed description about item-dependent transaction and the process carried out in this step could also be found in (Ventura, 2016). Then, each reducer processes a different item enabling a higher level of parallelism. The reducer creates an independent local FP-Tree and growths its conditional FP-tree recursively. The information contained in each node is a counter variable for each target (certificated, only explored, only viewed and only registered as it will be described in subsequent sections). The quality measures are calculated in this step since each node has enough information to do it, in our case: $Support_{target}$ and *Confidence*. Finally, taking the set of FP-Trees, the set of variables that form each subgroup is obtained by traversing each single FP-Tree.

---

**Algorithm 1** SD-EDM-DFPTree - Step 2.2 - Parallel counting
---

**function** mapperParallelComputing(instance)
 1: **for all** value in instance **do**
 2:   *emit*(*value,* 1)
 3: **end for**
**end function**
**function** reducer(element, Iterator<frequencies>)
 1: *globalFrequency* ← 0
 2: **for all** frequency in frequencies **do**
 3:   *globalFrequency + = frequency*
 4: **end for**
 5:   *F −list.append*(*element, globalFrequency*)
**end function**

---

**Algorithm 2** SD-EDM-DFPTree - Step 2.4 - Parallel FP-Growth

**function** mapperPFP(item, instance, F-list)
 1: *a[]* ← *split*(*instance*)
 2: **for** j = instance |- 1 to 0 **do**
 3:    *emit*(*a[j], a[0]..a[j]*)
 4: **end for**
**end function**
**function** reducer(item, Iterator<instances>)
 1: *LocalFpTree* ← *null*
 2: **for all** instance in instances **do**
 3:    *insert_build_fp_tree*(*LocalFpTree, instance*)
 4: **end for**
 5: *Mine LocalFpTree recursively*
**end function**

### 3.3. Post-processing

The final action of the proposed methodology is the post-processing step, which is responsible for filtering the rules in two different ways: 1) selecting the most interesting rules that satisfy all the courses, 2) erasing redundant rules.

**Step 3.1.** In order to select the most interesting and reliable rules that appear in most of the courses, two parameters have been used aiming at obtaining the best rules. First, the Confidence measure is used for quantifing the precision or reliability of the subgroups/rules. A threshold value of 0.8 has been set because this is the value most widely used in similar works (Ventura, 2016). Second, the proposed SD algorithm also includes the possibility of setting up a specific value regarding the number of courses in which a rule/subgroup appears. This filter is therefore useful to take those rules that are present in all the courses or only in a specific number of courses.

**Step 3.2.** Each reducer returns the best subgroups after mining its local FP-Tree to produce the subgroups. In this regard, the driver collects each result from each reducer, performing a global sorting by the quality measure. After that, because of the fact that the main drawback of using SD is the large number of similar subgroups that can be obtained, our proposal performs a post-procesing step to remove redundant rules. A rule is redundant (Zaki, 2004) if a more general rule having a higher or equal confidence exists. Thus the non-redundant rules are those that are most general, i.e., those having minimal antecedents and consequents, in terms of subset relation. A rule is more general if it has the same right-hand-side but one or more items removed from the left-hand-side. Formally, a rule $X \to Y$ is redundant if for some $X' \subset X$, $confidence(X' \to Y) \geq confidence(X \to Y)$. As a matter of clarification, let us consider two sample subgroups with the same value of Confidence (both are equally reliable) in the form of $Attribute_1 \to target_2$ and $Attribute_1 \wedge Attribute_3 \to target_2$. The first rule is a subset of the second one so, once the first one is obtained, the second rule can be removed. In this way, similar subgroups might be removed to promote those with smaller number of attributes, reducing the total number of subgroups and, therefore, easing the interpretation of the final results.

## 4. Experimental Analysis

The aim of this section is to describe the experimental analysis designed to validate the proposed methodology on a real-world scenario. In this regard, data as well as some pre-processing steps are first described. Then, the SD algorithm used in the proposed methodology is compared to traditional approaches for mining subgroups. Finally, some post-processing steps as well as a description of the obtained rules are also performed.

### 4.1. MOOC Data Description

The real-dataset[1] used in this work corresponds to de-identified data from the first year (Academic Year 2013: Fall 2012, Spring 2013, and Summer 2013) of 16 MITx and HarvardX courses on the edX platform (Harvardx, 2014). This is the first large scale public MOOC data repository that contains information (aggregated records) about diverse kinds of learning interactions and outcome features for the same students who took multiple courses on the platform.

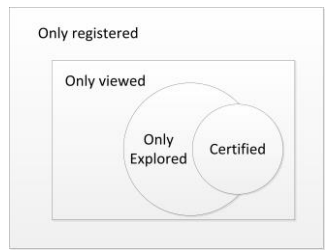

**Figure 3.** Type of students.

In this dataset, four subpopulations of interest (see Figure 3) within each course (Ho, 2014) were identified. Here, it is obtained that 43,196 registrants earned certificates of completion (Certified); 35,937 students explored half or more of course contents, including video and assessments but were not certified (Only Explored); 469,702 registrants viewed less than half of the content (Only Viewed); and 292,852 students never engaged with the online content (Only register). In total, there are 841,687 registrations from 597,692 unique users across the first year of HarvardX and MITx courses. During the de-identification process the original user ID or username, and the IP addresses where removed from the original file. The de-identification process also removes outliers and highly active users because these users are more likely to be unique and therefore easy to re-identify. Thus, the final number of rows after the de-identification was 641,138 from the original 841,687.

Figure 4 illustrates the total number of students in each of the 16 courses as well as the number of students in each of the four categories described above. About the number of students, most of the courses include information of a number of students that ranges between 20,000 and 60,000, with the exceptions of HarvardX/CS50x/2012, which includes about 170,000 students, and the courses MITx/2.01x/2013_Spring, MITx/3.091x/2013_Spring and MITx/8.MReV/2013_Summer that comprise less than 10,000 students. About the percentage of students in each category, Only Resgistered and Only Viewed are much larger than Only Explored and Certified students. The category that includes a largest number of students in almost all the courses is Only Viewed (with a percentage between 34% and 93%) followed by Only Registered (with a percentage between 4% and 62%). The two small categories with very similar number of students are Only Explored (with a percentage between 0,5% and 6%) and Certified (with a percentage between 0,7% and 8%).

---

[1]It was created on May 14, 2014. https://dataverse.harvard.edu/dataset.xhtml?persistentId=doi:10.7910/DVN/26147

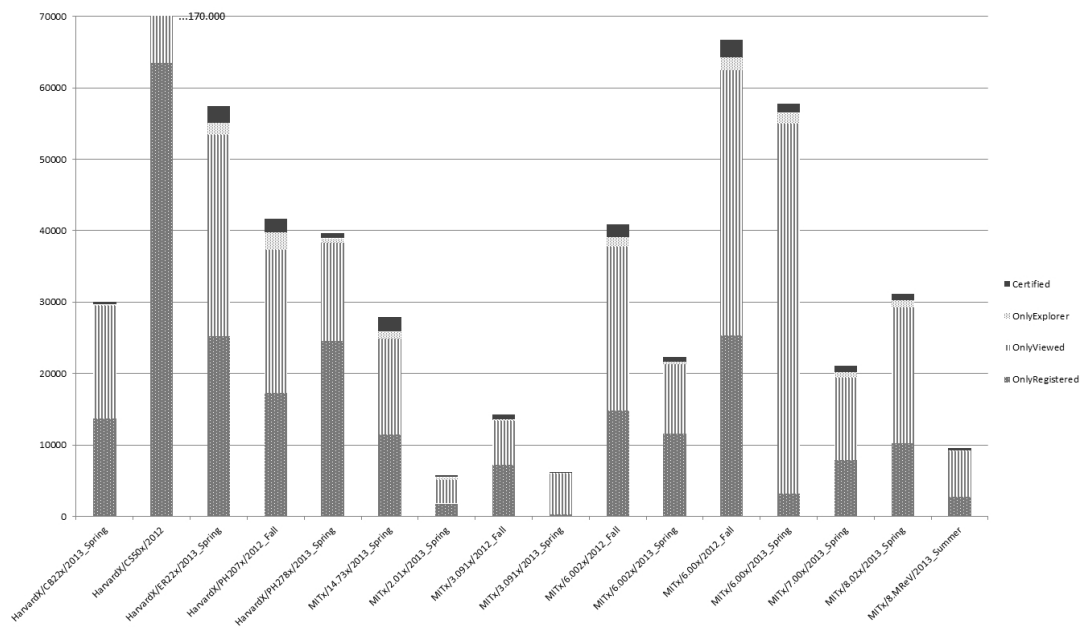

**Figure 4.** Number of students in each category by course.

In this dataset, each record represents the activity of a unique student in one edX course. It includes administrative information as well as information generated from users' interaction. Next, a detailed description of the variables is included:

- *course_id*. Institution (HarvardX or MITx), course name, and semester.
- *userid_DI*. First portion identifies dataset and second portion is a random ID number.
- *registered*. Determines whether the user is registered in a course.
- *viewed*. Anyone who accessed the "Courseware" tab (the home of the videos, problem sets, and exams) within the edX platform for the course.
- *explored*. Anyone who accessed at least half of the chapters in the course.
- *certified*. Anyone who earned a certificate. Certificates are based on course grades, and depending on the course, the cutoff for a certificate varies from 50% to 80%.
- *final_cc_cname_DI:*. Country name or corresponding continent/region name.
- *LoE*. Highest level of education completed: Doctorate, Master, Bachelor, Secondary and LessthanSecondary.
- *YoB*. Year of birth where the oldest date corresponds to the year 1931.
- *gender*. Possible values: m (male), f (female) and o (other).
- *grade*. Final grade in the course, ranges from 0 to 1.
- *start_time_DI*. Date of course registration.
- *last_event_DI*. Date of last interaction with course, blank if no interactions beyond registration.

- *nevents*. Number of interactions with the course, recorded in the tracking logs, blank if no interactions beyond registration and it ranges from 1 to 197757.
- *ndays_act*. Number of unique days student interacted with course, ranges from 1 to 205.
- *nplay_video*. Number of play video events within the course, ranges from 1 to 98517.
- *nchapters*. Number of chapters (within the Courseware) with which the student interacted, ranges from 1 to 48.
- *nforum_posts*. Number of posts to the Discussion Forum, ranges from 0 to 20.
- *roles*. Identifies staff and instructors.

As a matter of summarizing the information, all these attributes can be categorized into four main groups:

- *Identification attributes:* course_id and userid_DI.
- *Subgroup/Category attributes:* registered, viewed, explored and certified.
- *Personal/Demographic attributes*. final cc cname DI, LoE, YoB and gender
- *Activity/Engagement attributes*. grade, start time DI, last event DI, nevents, nchapters, ndays_act, nplay_video, nforum_posts and roles.

### 4.2. Preprocessing

Preprocessing is a really important step for obtaining interesting subgroups since despite the fact that some attributes do provide useful information, others require any kind of transformation. In this regard, the following attributes were removed from the dataset due to they do not provide any useful information (they are generally ids and dates):

- *userid_DI*. The final aim is not to describe specific cases but the general behavior so this attribute does not provide any useful information.
- *start_time_DI*. Since some courses started earlier than others, the comparison is not fair. Besides, it should be highlighted that no information about when a course is opened is provided so it cannot be used in our proposal.
- *last_event_DI*. Similarly to *start time DI*, each course starts and ends in a different date so the comparison is not fair. Besides, some MOOCs are available but it is not possible to obtain a certification so this attribute could add misleading information.
- *roles*. It does not provide any information since all the instances have the staff value.
- *registered*. This attribute denotes whether a user is registered within the course and, therefore, it includes a true value for all the users. Hence, this attribute does not provide any useful information and can be removed.

The following attributes are transformed into new ones:

- *registered*. We have created a new attribute named *onlyregistered* starting from registered users who are not Viewed, neither Explored nor Certified.
- *viewed*. This attribute denotes Viewed users who can be also Explored and Certified users. We have created a new attribute named *Onlyviewed* for Viewed users who neither are Explored nor certified users.
- *explored*. This attribute defines Explored users who can be also Certified users.

We have created a new attribute named *Onlyexplored* for Explored users who are not Certified users.

Additionally, since some attributes are defined in a continuous domain, they cannot be directly used in the SD proposal (only discrete attributes are considered since it is based on an exhaustive search methodology). Thus, all the continuous attributes were transformed into a discrete domain through the two following methods:

- A manual method for discretizing the attributes *YoB* and *grade* was carried out. This method consists on manually selecting the cut points for splitting the continuous values. In the case of the attribute *YoB* the cut points were: 1999, 1993, 1983, 1973 and 1963, and the corresponding to the labels: *<18*, *18-24*, *25-34*, *35-44*, *45-54*, *>54*. As for the *grade* attribute, only a single cut point was considered, that is, the value 0.5 to divide the grade into *low* and *high*.
- An automatic method for discretizing the attributes (*nevents*, *ndays_act*, *nchapters*, *nplay_video* and *nforum_posts*) was performed through the equal-width discretization algorithm, which is one of the most well-known unsupervised discretization (Dougherty, 1995) algorithms. In all the cases, three categories or labels were considered: *low*, *medium* and *high*. It is important to highlight that these attributes were discretized depending on the course since each one has its own nature and, therefore, the range of values may vary from a course to another. As a matter of clarification, some courses include a high number of videos, whereas other courses include a higher number of theoretic chapters. Thus, for a course that includes only 5 videos, a value of 4 played videos may be considered high, whereas for a course including 50 videos that same value (5 played videos) might be extremely low. Hence, with the aim of avoiding this problem, all the attributes were discretized with regard to each course.

Finally, it is important to highlight that the name of two attributes in the original dataset has been renamed to improve the comprehensibility. Thus, *final_cc_cname_DI* is now *contryName*, whereas *YoB* was renamed as *age*.

### *4.3. Post-processing*

As it was described in previous sections, one of the main disadvantages of using a traditional SD algorithm is the huge number of subgroups that is possible to be obtained, which hampers the interpretability from an expert's point of view. Besides, these subgroups can be highly overlapped so a large number of such subgroups do not provide additional information, that is, the very same information is repeated but expressed in a different way. As previously described, the proposed methodology tackles this problem with a post-processing step where interesting and not redundant rules are obtained.

In order to prove the usefulness of the post-processing step, the number of subgroups or interesting rules for the 16 courses is illustrated in Figure 5, considering either a post processing step or not (redundant rules are not removed). In this analysis, no interesting rule (having a Confidence value higher than 0.8) were obtained for the category Only Registered for any of the 16 courses. This issue is caused by considerable differences in demographics, which are described in the context of the diversity of course offerings, the intentions of the instructors teams, and the outreach and dissemination efforts of course teams.

Analysing the number of rules obtained for each target or category, it was obtained that in the worst scenario an improvement of one order of magnitude with regard to the number of subgroups was obtained (reducing from 66 Only Explored and 58 Certified rules to only 5 rules). On the contrary, in the best scenario, close to two orders of magnitude (from 781 Onlyviewed rules to only 11 rules) were achieved. Hence, in this case, the proposed methodology (considering a post-processing step) returned around 5 to 11 subgroups for each target. This number was enough to provide unknown information, but not so large to hamper the understanding process carried out by the end user. For instance, if the target OnlyViewed was considered, then the same model without a post-processing step would return more than 700 subgroups, which denotes highly overlapping and therefore a hampering of the final interpretability.

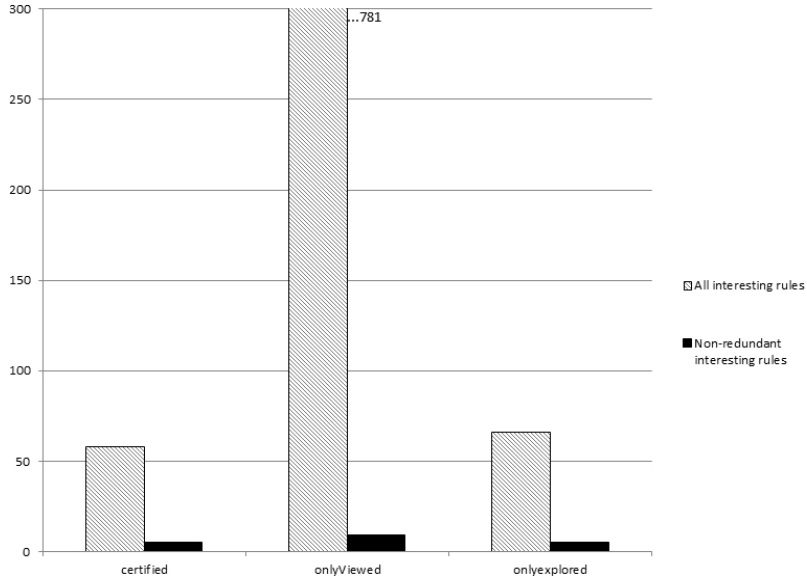

**Figure 5.** Number of subgroups obtained by the proposed approach with / without redundant rules for the different targets and 16 courses.

### *4.4. Runtime analysis*

The aim of this section is to analyse the runtime required by the proposed methodology when the proposed SD algorithm is performed in either a sequential way or by considering the MapReduce framework. It is important to highlight that both approaches return exactly the same subgroups and, therefore, there is no improvement in the final quality of the results (no study is required in this sense). The four categories or targets (Certified, Only Explored, Only Viewed and Only Registered) were considered and separately analysed (see Figure 6) since each of these subsets has a different distribution. According to the results shown in Figure 6, the implementation based on Spark (MapReduce) achieved the best performance (an almost constant runtime) regardless the target variable. When all the courses (a total of 16 courses) were considered at time, then the approach based on Spark outperformed the sequential model in one order of magnitude. It should be remarked that, in order to alleviate the possible desviations in the runtime, both implementations were run 30 times for each number of courses (from 1 to 16). Thus, results illustrated in Figure 6 are not absolute but average values.

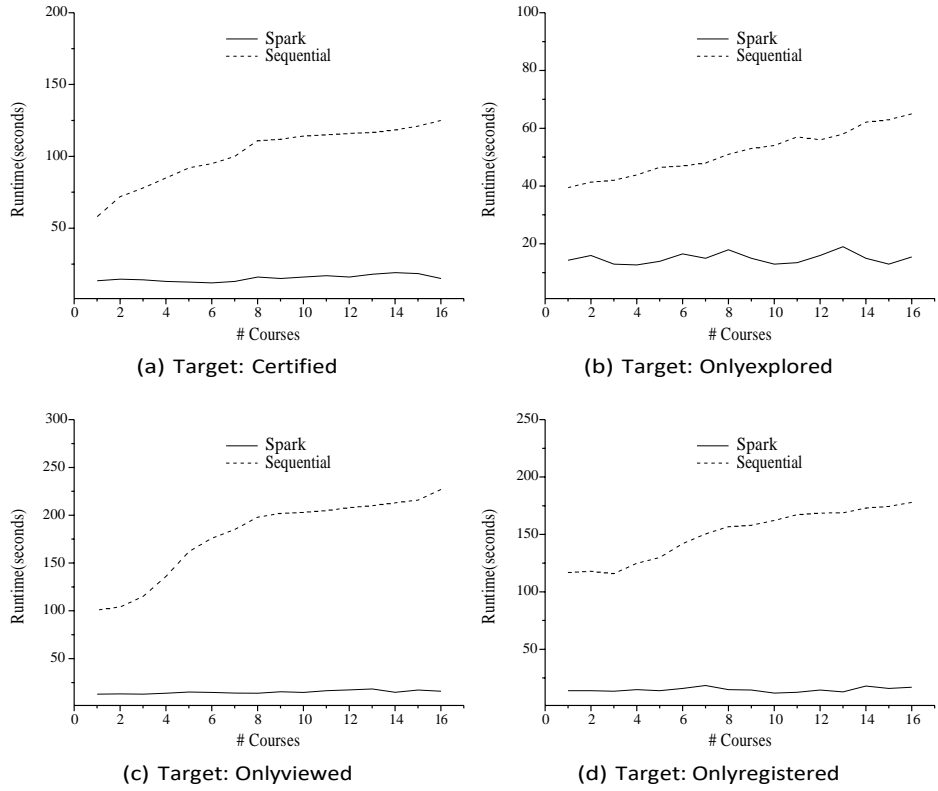

**Figure 6.** Runtime of the two versions (sequential and Spark) of the proposed SD algorithm.

## 4.5. Study of the resulting subgroups/rules

The goal of this section is to study the interestiness of the resulting set of rules/subgroups obtained by the proposed methodology. All these rules were properly obtained to match all the availables courses (16 MOOC), and these rules represents specific features for each of the categories or targets under study (Certified, Only-explorer, Onlyviewed and Onlyregistered). Each IF-THEN rule denotes a subgroup of students and two different measures are provided (Support within the target and Confidence). Support within the target quantifies the frequency of each set of features in its own category, thus it provides information of how frequent this implication is within the category. On the contrary, the confidence measure is used to evaluate the reliability of the implication denoted by the IF-THEN rules.

Table 2 shows the subgroups obtained for the Certified (anyone who earned a cer-tificate) category within the 16 courses. According to the results, it is obtained that only five interesting and not-redudant subgroups were obtained. All the obtained rules include three attributes in the antecedent, but two attributes always appear with high values: grade and NDayAct. Thus, these groups of certified students obtained a high grade, usually interacted with the course, and also match with another additional condition. The best subgroup (the one having the highest confidence value and a good specificity value) also includes the attribute NChapter with a high value, denoting that these certified students interacted with all or almost all the chapters of the course. There are two additional interesting rules (having a very similar support and a little lower confidence that the previous one) related to the categories Explored and Viewed. These rules represent two subgroups of certified students that also accessed at least half of the chapters (Explored) or accessed any time to the course (Viewed), respectively. Another rule with a similar confidence value describes Certified students who have a medium interaction with the course (Nevents=Medium). Finally, the last rule (the one with the lowest support value but including a good confidence value) describes Certified students that also have secondary as highest level of education.

| Rule | $Support_{target}$ | Confidence |
|---|---|---|
| **IF** grade=High **AND** NDaysAct=High **AND** NChapter=High **THEN** certified=True | 0.37 | 0.87 |
| **IF** grade=High **AND** NDaysAct=High **AND** explored=True **THEN** certified=True | 0.37 | 0.84 |
| **IF** grade=High **AND** NDaysAct=High **AND** viewed=True **THEN** certified=True | 0.37 | 0.82 |
| **IF** grade=High **AND** NDaysAct=High **AND** NEvents=Medium **THEN** certified=True | 0.32 | 0.82 |
| **IF** grade=High **AND** NDaysAct=High **AND** LoE=Secondary **THEN** certified=True | 0.06 | 0.80 |

**Table 2.** Rules obtained for the certified class/target.

| Rule | Support_target | Confidence |
|---|---|---|
| **IF** certified=False **AND** NDaysAct=Medium **THEN** onlyexplored=True | 0.30 | 0,99 |
| **IF** certified=False **AND** NChapters=Medium **THEN** onlyexplored=True | 0.18 | 0.99 |
| **IF** certified=False **AND** NPlayVideo=Medium **THEN** onlyexplored=True | 0.30 | 0.98 |
| **IF** certified=False **AND** NumberOfPosts=Medium **THEN** onlyViewed=True | 0.31 | 0.91 |
| **IF** certified=False **AND** NEvents=Medium **THEN** onlyexplored=True | 0.31 | 0.91 |

**Table 3.** Rules obtained for the onlyexplored class/target.

Table 3 shows five interesting and not-redundant subgroups for Only Explored learners. They have two attributes in the antecedent and the first one is always the same attribute (certificate=False), denoting that all these students did not obtained a certificate. All the obtained rules have a high confidence value (greather than 0.90) even when the support value highly differ from a rule to another. The best subgroup (the one with the highest confidence value) describes that Only Explored students interacted an average number of days with the course (NDaysAct=Medium). The second best rule illustrates that Only Explored students interacted with an average number of chapters of the course (NChapter=Medium). The third rule is related to an average Number of play video events (NPlayVideo=Medium). The next rule shows that Only Explored students usually send an average number of posts to discussion forums (NumberOfPosts=Medium). Finally, the last rule determines that Only Explorer students have a average number of interactions with the course (Nevents=Medium).

According to the results shown in Table 4, there are eleven interesting and not-redundant subgroups for the Only Viewed category or class. All these rules include two attributes in the antecedent and one attribute (denoting students who obtained a low final grade) always appears. All the obtained rules have a high confidence, whereas the support values highly differ from rule to rule (the values vary from 0.99 to 0.02). The best rule has a confidence value close to 1 and a really good support value, and it describes that Only Viewed students interacted with only few of chapters of the course (NChapter=Low). The second best rule shows that Only Viewed students usually send a low number of post to the discussion forums (NumberOfPosts=Low). The third rule denotes that Only Viewed students interacted few days with the course (NDaysAct=Low). The fourth and fifth rules (having similar confidence but lower support values than the previous one) show that Only Viewed students may be women or men. Then, the next five rules (having similar confidence, but a lower support value) show that Only Viewed students have completed a different level of education. The largest subgroup is related to bachelor studies, following by Secondary, Master and, finally, both LessthanSecondary and Doctorate represent the smallest groups. The last rule determines that Only Viewed students have a low number of interactions with the course (Nevents=Low).

| Rule | Support_target | Confidence |
|---|---|---|
| **IF** grade=Low **AND** NChapter=Low **THEN** onlyViewed=True | 0.55 | 0.99 |
| **IF** grade=Low **AND** NumberOfPosts=Low **THEN** onlyViewed=True | 0.99 | 0.93 |
| **IF** grade=Low **AND** NDaysAct=Low **THEN** onlyViewed=True | 0.73 | 0.93 |
| **IF** grade=Low **AND** gender=f **THEN** onlyViewed=True | 0.28 | 0.92 |
| **IF** grade=Low **AND** gender=m **THEN** onlyViewed=True | 0.12 | 0.92 |
| **IF** grade=Low **AND** LoE=Secondary **THEN** onlyViewed=True | 0.26 | 0.94 |
| **IF** grade=Low **AND** LoE=Bachelors **THEN** onlyViewed=True | 0.34 | 0.93 |
| **IF** grade=Low **AND** LoE=Masters **THEN** onlyViewed=True | 0.18 | 0.93 |
| **IF** grade=Low **AND** LoE=LessthanSecondary **THEN** onlyViewed=True | 0.02 | 0.93 |
| **IF** grade=Low **AND** LoE=Doctorate **THEN** onlyViewed=True | 0.02 | 0.92 |
| **IF** grade=Low **AND** NEvents=Low **THEN** onlyViewed=True | 0.67 | 0.92 |

**Table 4.** Rules obtained for the only viewed class/target.

| Rule | Support_target | Confidence |
|---|---|---|
| **IF** countryName=Unknown/Other **AND** gender=m **THEN** onlyregistered=True | 0.07 | 0.92 |
| **IF** countryName=Unknown/Other **AND** age=25-34 **THEN** onlyregistered=True | 0.06 | 0.92 |
| **IF** countryName=Unknown/Other **AND** LoE=Bachelor's **THEN** onlyregistered=True | 0.04 | 0.92 |

**Table 5.** Rules obtained for the only registered class/target.

Finally, considering the Only Registered category, we have considered only attributes related to personal/demographic information (countryName, LoE, age and gender) because this type of users have no interaction with the course. It is remarkable that no interesting subgroup was found for any of the courses. Hence, Table 5 includes interesting and non-redundant rules that match 9 courses (from the total of 16 courses). As it is illustrated in Table 5, there are only three interesting and not-redundant rules that match with 9 courses from the total of 16 available courses. These rules have 2 attributes in the antecedent and one of these attributes is related to the country of the learners. The three rules have the same high confidence value (0.92) but a very low support value (it ranges from 0.04 to 0.07). The first rule describes that Only Registered students are usually male for those countries that are not in the top. The next rule describes learners with an age that ranges between 25 and 34, so it determines a second subgroup of Only Registered students who are middle age learners and who belong to a country that is not in the top. Finally, the last rule shows learners that have Bachelor studies as the highest level of education, and who belong to a country that is not in the top.

The information provided by all the previous SD rules (see Tables 2, 3, 4 and 5 can be used for different purposes: showing a general description of categories of learners in MOOC; predicting these type of learners in new courses; future recomendations as well as personalization purposes. All these discovered rules provided a general description of the different types of learners in MOOCs. Additionally, these SD rules can also be used as classifcation rules to group new students into these subgroups. Finally, the extracted information can be used for recomending learners from different categories what they should do (actions, activities, etc.) in order to change their category within the MOOC. For each class/target we have different rules or subgroups that describe these learners. So, using this information we can recommend for example what must to do a only viewer learners to can go over explored learners, or a explored learners to can go over a certified learners in a similar way than other previous recommendation MOOCs approaches (Agrawal, 2015) (Bousbahi, 2015) Hajri (2017). Finally, these SD rules can be also used for content personalization purpose in the same way than previous MOOC personalization approaches (Brinton, 2015; McLoughlin, 2013). In our case, firstly, it is necessary to determine/find for each discovered subgroup what are the most accessed chapters, viewed videos, done quizzes, etc. Then, after classifying new learners into subgroups, we can recommend to these learners for example to visit the most viewed video of previous learners in this subgroup. It is important to notice that for the Only registered target/class we have not obtained any interesting subgroups that appear in all the courses. So, we have tried to find subgroups that appear in less number of courses and we only discovered some subgroups when we reduced the number of courses to 9. However, these subgroups are very small and does not provide much information about only registered users. They are all users from an unknown or other than the top 35 country. There are several small subgroups of these users which some of them are men, others have a age between 25-34 year, and others studied bachelors. In the future, we want to apply our SD proposals in a more number of MOOC datasets in order to test if we can discover some interesting subgroups about the registered target/class and if we discover the same subgroups of the others targets/classes (certified, only explorerer and only viewer).

Finally, we have taken the largest courses (HarvardX/CS50x/2012, HarvardX/ER22x/2013_Spring, MITx/6,00x/2012_Fall, MITx/6,00x/2013_Spring) comprising more than 55,000 students each and we have analysed, through a parallel graph (multi-criteria decision making graph) how the average results for each metric ($Support_{target}$ and *Confidence*) behave for each type of learner and each course. As it is summarized in Figure 7, the average reliability is almost constant for each dataset and each type of learner. In fact, the average values range from 0.9 to 1.0. On the contrary, analysing the Support relative to each target, it is obtained that the average values highly differ from learner to learner and it is a matter of the variability of the groups of learners so the higher this variability in a group, the more difficult is to find features for this group. Nevertheless, the results are quite similar from one dataset to another meaning that the differences fall on the type of learners and not on the dataset.

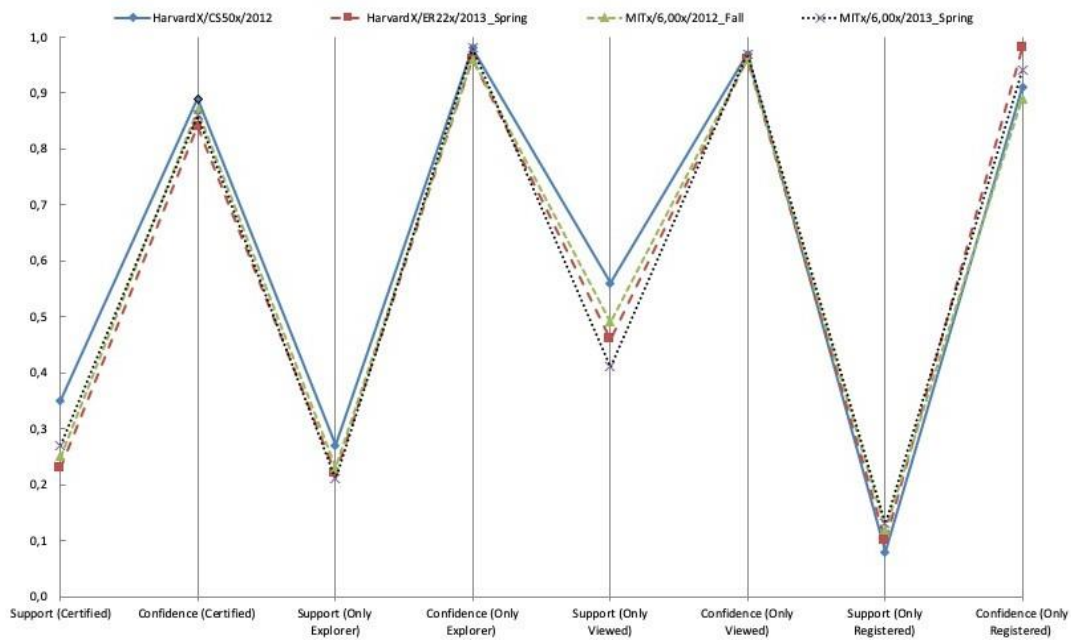

**Figure 7.** Summary of the results for the 4 largest courses, considering each metric ($Support_{target}$ and *Confidence*) and each type of learner.

## 5. Conclusions

In this paper, we have proposed a subgroup discovery algorithm for describing different types of learners in MOOCs. The proposed methodology is based on MapReduce so interesting subgroups can be described on really large datasets. It includes an scalable method for mining interesting and non-redundant subgroups in the form of IF-THEN rules. In the experiemental stage, data gathered from a MOOC were analysed, considering the first year of 16 MITx and HarvardX courses on the edX platform. Results showed that the proposed MapReduce approach outperformed traditional sequential subgroup discovery approaches, achiving a runtime that is almost constant. In order to discover the most interesting rules that appear in most of the courses under study, the proposed methodology includes two specific threshold values related to the minimum reliability of the rules as well as the number of courses in which each subgroup must appear. In this way, our approach is able to discover the best not-redundant rules that appear in all the courses or in a number of courses.

This work have some restrictions related to the subgroup discovery task and the data to be analysed. Firstly, subgroup discovery requires a class or target attribute. In this regard, the proposal cannot be directly apply to any MOOC dataset in an automatic way and it is therefore required that the dataset provides not only the traditional descriptive atributes about the learners but also some class or target attributes. In our case, this step was not necessary since MITx and HarvadX datasets already provided us with such information. Another restriction of this work is related to the structure of the MOOC data. In this paper we have used a summary dataset with aggregated data about all the courses that were gathered by MITx and HarvardX Dataverse. This summary data provided us with a coarse set of features that can be useful for consistently identifying very high level patterns of engagements across different courses. However, some of the previous related works about discovering learners engagement used temporal information instead of summary information, that is, they used time variant variables and they looked for trends about how that variables changed over time (Kizilcec, 2013). Authors used different granules of time for assigning labels such as days, weeks, moths, etc. The fine-grained time slices can be very precise but they can be more specific to a course. In the future, we want to adapt the porposed algorithm for using not only summary data but also temporal information in order to discover subgroups at different level of granularity.

**Acknowledgement(s)**

This research was supported by the Spanish Ministry of Economy and Competitiveness and the European Regional Development Fund, projects TIN2017-83445-P.